# Theoretical analysis of the force on the end face of a nano-filament exerted by an outgoing light pulse

Masud Mansuripur[†] and Armis R. Zakharian[‡]

[†]College of Optical Sciences, The University of Arizona, Tucson, Arizona 85721 <masud@optics.arizona.edu>
[‡]Corning Incorporated, Science and Technology Division, Corning, New York 14831

**Abstract.** The slight deformations observed upon transmission of a light pulse through a short length of a silica glass nano-filament offer the possibility of determining the momentum of light inside the filament. Using precise numerical calculations that take into account not only the electromagnetic momentum inside and outside the filament, but also the Lorentz force exerted by a light pulse in its *entire* path through the nano-waveguide, we conclude that the net effect of a short pulse exiting the nano-filament should be a pull force on the end face of the filament.

**1. Introduction**. In a recent paper [1], W. She, J. Yu and R. Feng reported the slight deformations observed upon transmission of a light pulse through a short length of a silica glass nano-filament. Relating the shape and magnitude of these deformations to the momentum of the light pulse inside and outside the filament, these authors concluded that, within the fiber, the photons carry the Abraham momentum. In our view, the interpretation of the experimental results of She *et al* requires precise numerical calculations that properly account for the electromagnetic (EM) momentum inside and outside the fiber, as well as for the Lorentz force exerted on the fiber by the light pulse in its *entire* path through this nano-scale waveguide [2-4]. Our goal in the present paper is to provide a theoretical analysis of force and momentum both inside and in the surrounding regions of the nano-fiber, when a short (femtosecond) light pulse approaches the end face from within, then exits into the vacuum outside.

The momentum of a light pulse in vacuum is given by $\mathcal{E}_{\text{pulse}}/c$, the ratio of the pulse energy to the speed of light in vacuum, only when the pulse has a cross-sectional diameter (in the *xy*-plane perpendicular to the propagation direction *z*) that is much greater than the light's wavelength $\lambda$. In other words, $\boldsymbol{p} = (\mathcal{E}_{\text{pulse}}/c)\hat{\boldsymbol{z}}$ is valid only when there is negligible diffraction broadening during propagation of the beam along the *z*-axis. The diameter of the light spot emerging from a nano-fiber is typically ~500 nm, which is comparable to a wavelength of the visible light. Consequently, the emergent momentum along the axis of the fiber (*z*-axis) is much reduced compared to $\mathcal{E}_{\text{pulse}}/c$. This problem will be addressed in Section 2.

A potential source of error when interpreting experimental results is the use of the *phase* refractive index $n_\phi$ of the filament material to determine the Abraham momentum inside the nano-waveguide; the correct parameter in this situation is the *group* refractive index $n_g$. In the case of an ordinary glass slab, one might be able to ignore dispersion effects and assume dispersionless propagation, in which case $n_\phi$ and $n_g$ will be nearly the same. The validity of this "dispersionless" approximation, however, is no longer obvious when one deals with propagation within a waveguide, especially when the filament has been pulled so drastically (as in the reported experiments) as to force a substantial fraction of the optical energy into the air (or vacuum) surrounding the nano-fiber. Our computer simulations reported in Section 3 reveal that the group index of a nano-filament made of a dispersionless material having $n_\phi = 1.5$ is $n_g = 1.619$ (filament diameter = 460 nm, pulse duration ~ 10 fs, central wavelength $\lambda_o = 0.6 \, \mu\text{m}$). Using $n_\phi$ instead of $n_g$ in our example would result in an 8% over-estimation of the Abraham momentum inside the fiber.

The Abraham momentum is only one component of the momentum of light inside a dielectric, namely, the electromagnetic component $\boldsymbol{p}_{EM}$; the other component is mechanical, denoted by $\boldsymbol{p}_{mech}$. (Mechanical momentum is *not* the same thing as the difference between the Minkowski and Abraham momenta, as some authors have suggested; it is half as much under certain circumstances [6]; for a thorough review of the subject see [7].) A correct accounting for the deformation of the nano-filament requires the balancing of $\boldsymbol{p}_{EM}+\boldsymbol{p}_{mech}$ inside the fiber against the pure EM momentum which emerges from the fiber into the free space. Our numerical simulations in Section 3 provide an estimate for the mechanical momenta of the incident and reflected light pulses, showing $\boldsymbol{p}_{mech}$ to be comparable to $\boldsymbol{p}_{EM}$ and, therefore, impossible to neglect. In fact, including the contribution of $\boldsymbol{p}_{mech}$ in the analysis will reverse the direction of the force exerted on the nano-filament, resulting in a net pull (rather than push) force.

In Section 4 we summarize our results and point out the limitations of our method of calculating $\boldsymbol{p}_{mech}$, which arise from neglecting the elastic properties of the nano-fiber and ignoring the possibility of acoustic wave generation and propagation. While the general idea of monitoring the mechanical response of a nano-filament to the passage of a light pulse is meritorious and could potentially provide answers to fundamental questions pertaining to the momentum of light, we believe careful analysis and more accurate measurements are needed before the Abraham-Minkowski controversy can be settled.

**2. Momentum of light emerging from a nano-filament into the free space**. In the original Einstein box thought experiment [5], which involves an empty box on a frictionless rail, with light emerging from the wall on the left, traveling the length of the box, then impinging on a perfect absorber on the right-hand side, one must recognize that, if the pulse emerging from the wall on the left has a small cross-sectional diameter (e.g., emanating from a nano-fiber), it will expand, due to diffraction, into a spherical wave as it propagates to the right. One can then readily see that the pulse's momentum must be less than $\mathcal{E}_{pulse}/c$; the smaller the transverse dimensions of the light source, the greater will be the deviation of the pulse's momentum from $\mathcal{E}_{pulse}/c$.

Figure 1 shows a special Einstein box designed for the light emanating from the tip of a nano-fiber located on the left-hand side. An absorbing layer is applied uniformly over a hemi-spherical surface of radius $R$ on the right-hand side. Suppose a short pulse of energy $\mathcal{E}_{pulse}$ and momentum $\boldsymbol{p}=p_z\hat{\boldsymbol{z}}$ emerges from the fiber and travels for a duration $\tau=R/c$ before getting fully absorbed. At first, the box acquires a momentum of $-\boldsymbol{p}$ and moves to the left with a velocity of $p/M$, where $M$ is the mass of the box (assumed to be very large). The box comes to a halt after traveling a distance $\Delta z_{box}=-(p/M)(R/c)$. During the same time interval, the light pulse moves to the right (while spreading by diffraction) and transfers its mass of $\mathcal{E}_{pulse}/c^2$ to the hemi-spherical surface. To be specific, let us assume that the fiber-tip acts as a point dipole oscillating along the $x$-axis. The electromagnetic $E$- and $H$-fields arriving on the hemi-spherical surface will have amplitudes $(E_o\hat{\boldsymbol{\theta}}/R)\sin\theta$ and $(H_o\hat{\boldsymbol{\phi}}/R)\sin\theta$, resulting in an integrated intensity

$$½\int_0^{\pi/2} 2\pi R^2 \cos\theta |\boldsymbol{E}\times\boldsymbol{H}| \mathrm{d}\theta = \pi E_o H_o \int_0^{\pi/2} \cos\theta \sin^2\theta \mathrm{d}\theta = ⅓\pi E_o H_o. \qquad (1)$$

Thus, upon absorption, the center of mass of the light pulse will have moved to the right by

$$\Delta z_{light} = \frac{\int_0^{\pi/2} 2\pi R^2 \cos\theta (R\sin\theta)|\boldsymbol{E}\times\boldsymbol{H}|\mathrm{d}\theta}{\int_0^{\pi/2} 2\pi R^2 \cos\theta |\boldsymbol{E}\times\boldsymbol{H}|\mathrm{d}\theta} = 3R/4. \tag{2}$$

In the absence of external forces, however, the center of mass of the system cannot have moved in the process. Setting the net displacement of the center of mass equal to zero yields the momentum of the light pulse as $\boldsymbol{p} = (3\mathcal{E}_{pulse}/4c)\hat{\boldsymbol{z}}$. In other words, the light exiting the fiber carries only 75% of the momentum that is frequently (albeit erroneously) assigned to it.

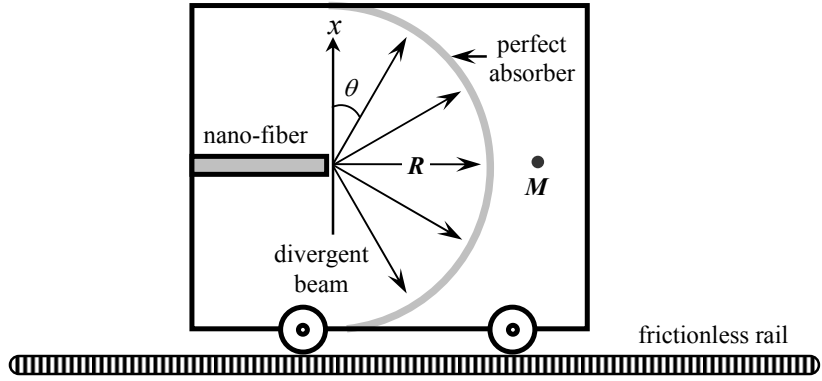

**Figure 1**. Einstein box on a frictionless rail. A short pulse of light having energy $\mathcal{E}_{pulse}$ and momentum $\boldsymbol{p} = p_z\hat{\boldsymbol{z}}$ emanates from the tip of a nano-fiber, then propagates toward a perfect absorber coated on a hemi-spherical surface of radius $R$. The mass of the box is $M$, the fiber tip coincides with the hemi-sphere's center, and the propagation time is $\tau = R/c$.

An alternative argument for the reduced momentum of a light pulse whose transverse dimensions are comparable to the wavelength $\lambda$ involves expressing the pulse's $E$- and $H$-field amplitudes as Fourier integrals in the $(\boldsymbol{k},\omega)$-space, followed by computing the Poynting vector, energy content, and momentum of the entire pulse; see the Appendix for details.

**3. Numerical simulations**. We describe the results of Finite Difference Time Domain (FDTD) numerical simulations for a short pulse of light traveling through and emerging from a silica glass nano-fiber. Figure 2(a) shows the cross-sectional profile of the simulated nano-fiber having a circular cross-section with a diameter $d = 460$ nm and refractive index $n_\phi = 1.5$. The assumed material dispersion for this filament is zero, that is, $\mathrm{d}n_\phi(\omega)/\mathrm{d}\omega = 0$, although guided-mode dispersion is automatically accounted for in our FDTD simulations. To facilitate the coupling of incident light into the filament, the medium of incidence at the top of Fig. 2(a) is assumed to have the same refractive index as the fiber. We launch into the nano-fiber a 10 fs Gaussian light pulse of central (vacuum) wavelength $\lambda_0 = 0.6$ μm. Once the pulse settles into a guided mode, its $z$-component of the Poynting vector $\boldsymbol{S}(\boldsymbol{r},t)$, integrated over the entire cross-sectional $xy$-plane, is monitored as a function of time. Plots of $\iint_{-\infty}^{\infty} S_z(x,y,z_0,t)\mathrm{d}x\mathrm{d}y$ versus time depicted in Fig. 2(b) show the optical energy flux at three different points along the $z$-axis, located at $z_0 = 2.5$ μm, $1.5$ μm, and $-0.5$ μm; also shown is the envelope of the pulse in each case. Over these short distances, the pulse propagates along the fiber's axis without any apparent distortion due to dispersion and/or attenuation. The total pulse energy is found to be 2.78 nJ, while the group refractive index (obtained by monitoring the pulse envelope's peak position versus time) turns out to be $n_g = 1.619$.

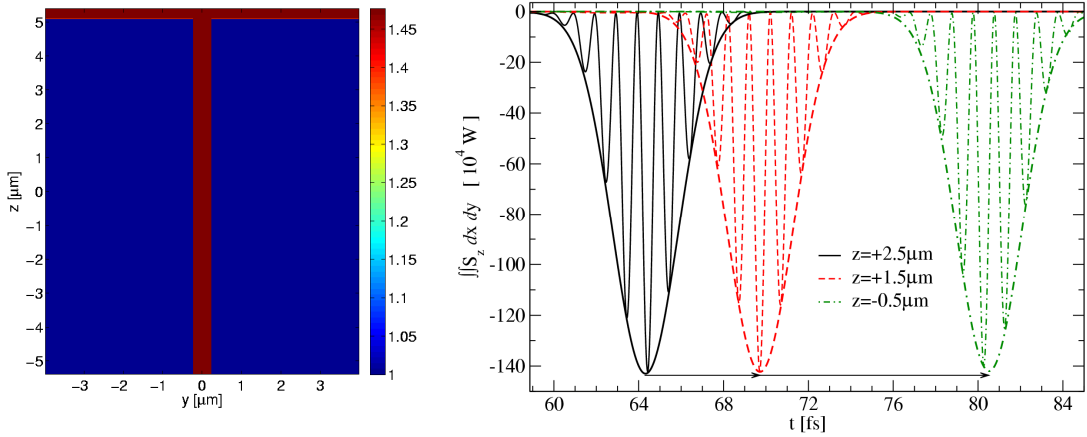

**Figure 2** (color online). (a) Cross-sectional profile of the simulated nano-filament: diameter $d = 460$ nm, phase refractive index $n_\phi = 1.5$. To facilitate the coupling of the incident pulse, the incidence medium at the top has the same refractive index as the fiber. (b) A 10 fs Gaussian light pulse of central wavelength $\lambda_o = 0.6\,\mu$m is launched into the fiber. Once the pulse settles into a guided mode, its integrated $S_z$ over the cross-sectional $xy$-plane (including the evanescent field) is monitored as a function of time. The solid (black), dashed (red), and dash-dotted (green) curves represent the rate of flow of optical energy at $z = 2.5\,\mu$m, $1.5\,\mu$m, and $-0.5\,\mu$m, respectively; also shown is the pulse's envelope in each case. The energy content of the pulse, obtained by integrating the flux over the pulse duration, is 2.78 nJ.

Figure 3 shows a computed graph of $\iint f_z(x,y,z,t_o)\mathrm{d}x\mathrm{d}y$, the integrated Lorentz force density over the cross-sectional $xy$-plane of the fiber, plotted versus $z$ at some fixed instant of time $t_o$. The Lorentz force density [2] is given by

$$\boldsymbol{f}(\boldsymbol{r},t) = (\boldsymbol{P}\cdot\nabla)\boldsymbol{E} + (\partial\boldsymbol{P}/\partial t)\times\mu_o\boldsymbol{H}, \qquad (3)$$

where $\boldsymbol{E}(\boldsymbol{r},t)$ and $\boldsymbol{H}(\boldsymbol{r},t)$ are the usual electric and magnetic fields, $\boldsymbol{P}(\boldsymbol{r},t) = \varepsilon_o(\varepsilon - 1)\boldsymbol{E}(\boldsymbol{r},t)$ is the polarization density within the nano-fiber, $\varepsilon_o$ and $\mu_o$ are the permittivity and permeability of free space, and $\varepsilon = n_\phi^2$ is the dielectric constant of the material. Let us assume that this force profile has been propagating (and will continue to propagate), *without* distortion or attenuation, at the constant group velocity of $c/n_g$ along the negative $z$-axis. Each volume element $\Delta v = \mathrm{d}x\mathrm{d}y\mathrm{d}z$ of the filament will thus experience the instantaneous longitudinal force $f_z(x,y,z + ct/n_g)\Delta v$. The time-integrated force being the mechanical momentum $\Delta p_z$ acquired by $\Delta v$ at an arbitrary instant of time, say, $t_o = 0$, the acquired mechanical momentum of the volume element $\Delta v$ will be

$$\Delta p_z^{(\mathrm{mech})} = \Delta v \int_{-\infty}^{0} f_z(x,y,z + ct/n_g)\mathrm{d}t = \Delta v(n_g/c)\int_{-\infty}^{z} f_z(x,y,z')\mathrm{d}z'. \qquad (4)$$

Integrating over the entire volume of the nano-fiber, we find the following expression for its total mechanical momentum at any instant of time:

$$p_z^{(\mathrm{mech})} = (n_g/c)\iiint_{-\infty}^{\infty}\mathrm{d}x\mathrm{d}y\mathrm{d}z\int_{-\infty}^{z} f_z(x,y,z')\mathrm{d}z'. \qquad (5)$$

The dashed (red) curve in Fig. 3, a plot of $\iint_{-\infty}^{\infty}\mathrm{d}x\mathrm{d}y\int_{-\infty}^{z} f_z(x,y,z')\mathrm{d}z'$ versus $z$, clearly indicates that the material elements within the light pulse have a mechanical momentum directed along the negative $z$-axis. At the leading edge (i.e., left-hand side of the figure), the total force integrated from the beginning to the mid-point of the pulse is $F_z = -1.88\times 10^{-3}$ N, with the minus sign signifying the direction of the force along the negative $z$-axis, which is the direction of propagation. The total force on the trailing edge (i.e., right-hand side of the figure), integrated from the mid-point to the end of the pulse, is $F_z = 1.88 \times 10^{-3}$ N. Thus while the leading edge pushes the host material's molecules forward, the trailing edge deprives them of their acquired mechanical momentum by applying a braking force. In accordance with Eq. (5), the area under the dashed (red) curve in Fig. 3 multiplied by $n_g/c$ must be the total mechanical momentum of the light pulse, $p_z^{(\text{mech})} = -3.3 \times 10^{-18}$ kg·m/s.

When the light pulse arrives at the exit facet, its mechanical momentum must be delivered to the fiber tip because, aside from a small fraction that returns with the reflected pulse, the incident mechanical momentum simply has nowhere else to go. When interpreting the results of nano-filament experiments, we believe the contribution of this mechanical momentum (delivered by the light pulse to the fiber's tip) should be taken into account. We are cognizant, of course, of the fact that the light pulse is followed in its path by an acoustic wave generated at the mount (to which the nano-fiber is affixed). For a rigid mount, this acoustic wave tends to essentially restore to their initial positions all the glass molecules that have been displaced forward by the light pulse. However, the acoustic wave arrives at the fiber tip sometime after the light pulse, as the latter typically travels much faster than the former. (The delay is proportional to the length of the fiber between the mount and the exit facet.) A complete analysis of the deformation dynamics of the fiber must, therefore, take into account not only the electromagnetic and mechanical momenta carried along by the light pulse, but also the acoustic interactions between the mount and the body of the fiber.

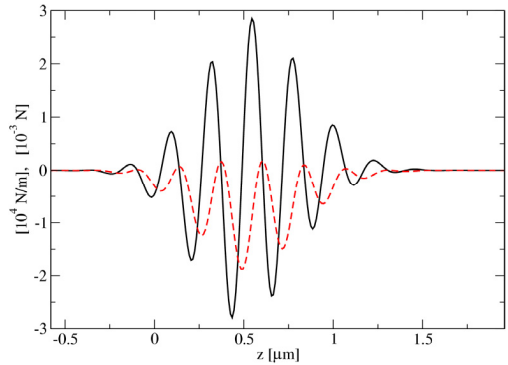

**Figure 3** (color online). At a fixed instant of time $t_o$, the solid (black) curve shows the integrated Lorentz force density over the *xy* cross-sectional plane of the nano-fiber versus *z* for the entire length of the pulse. The dashed (red) curve is the integral of the solid (black) curve along the length of the fiber from $z = -\infty$ to $z$. The total mechanical momentum at any given instant of time, obtained by multiplying the area under the dashed (red) curve with $n_g/c$, is $p_z = -3.3 \times 10^{-18}$ kg·m/s.

Figure 4 shows computed profiles of $E_x(x,y,z,t)$, the *x*-component of the *E*-field of the light pulse, at different instants of time in both *xz* and *yz* cross-sectional planes. The incident beam, being linearly-polarized along the *x*-axis, is responsible for the slight differences between the *xz* and *yz* cross-sectional profiles. The guided mode consists of propagating fields within the fiber as well as evanescent fields in the free-space region surrounding the fiber. In this and the following simulations, the fiber depicted in Fig. 2(a) was truncated at $z = 0.5$ μm to allow the light to emerge into the free space region below the fiber's tip. Once the pulse reaches the exit facet at $z = 0.5$ μm, it emerges into the free-space below the fiber, then proceeds to expand laterally via ordinary diffraction. At the same time, a small fraction of the incident pulse (~4.7% of the incident optical energy), bounces back and returns along the positive *z*-axis as a guided mode within the fiber.

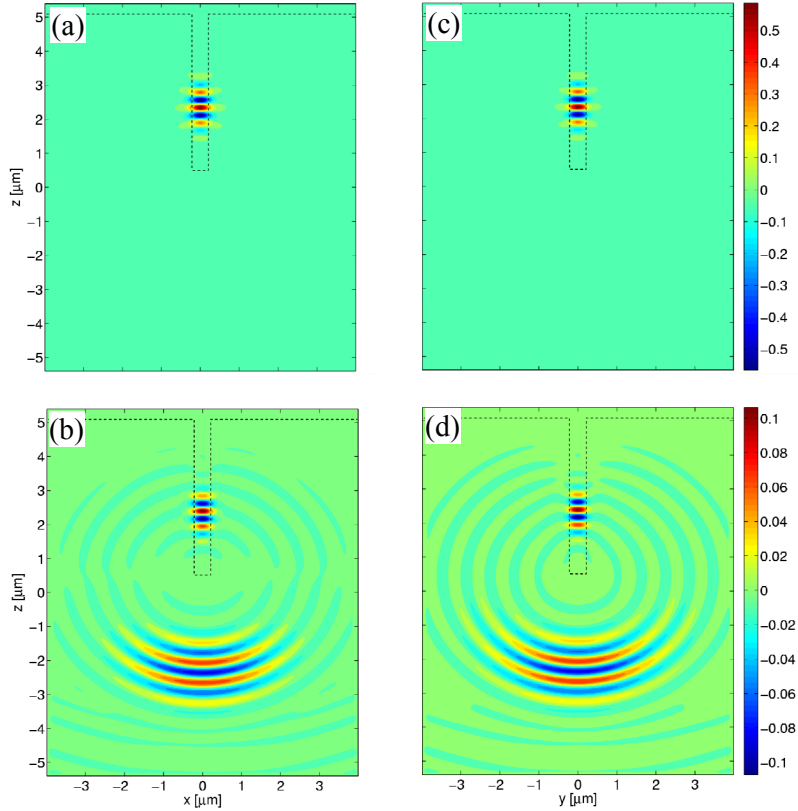

**Figure 4** (color online). Profiles of $E_x(x,y,z,t)$ at two instants of time in (a, b) $xz$ and (c, d) $yz$ cross-sectional planes. The incident beam is linearly-polarized along the $x$-axis, hence the slight differences between the $xz$ and $yz$ cross-sectional pulse profiles. Top row (a, c): at $t=65$ fs the pulse is moving down within the fiber. Bottom row (b, d): at $t=85$ fs a weak reflected pulse is moving up along the fiber's axis, while the transmitted light propagates downward and spreads laterally via diffraction. Note that the color-scale for (a, c) differs from that for (b, d). The guided mode consists of propagating fields within the fiber as well as evanescent fields in the surrounding free-space. Upon arriving at the exit facet ($z=0.5\,\mu$m), the pulse emerges into the free-space below and proceeds to expand laterally via ordinary diffraction. A small fraction of the incident pulse, carrying $\sim 4.7\%$ of its optical energy, bounces back and returns (within the fiber) along the positive $z$-axis.

Figure 5 shows plots of the energy flux, $\iint_{-\infty}^{\infty} S_z(x,y,z_0,t)\,dxdy$, versus time far above the exit facet at $z_0 = 2.5\,\mu$m (solid black), and also just below the exit facet at $z_0 = 0.48\,\mu$m (dashed green). The reflected pulse shows up in the solid black curve with $\sim 21$ fs delay due to the round-trip between $z=2.5\,\mu$m and the exit facet located at $z=0.5\,\mu$m. The transmitted and reflected optical energies in Fig. 5 add up to $\sim 98\%$ of the incident energy, with the remaining 2% also exiting the fiber, but propagating more or less radially away from the fiber's tip; this small fraction of the light cannot be captured by (numerical) monitors that are placed in cross-sectional planes perpendicular to the $z$-axis.

Figure 6 shows the computed Lorentz force exerted on the fiber as a function of time, $F_z(t) = \iiint_{-\infty}^{\infty} f_z(\mathbf{r},t)\,dxdydz$. The pulse reaches the tip at $t \approx 70$ fs and leaves the fiber by $t \approx 80$ fs. The force $F_z(t)$ experienced by the fiber during this interval oscillates between positive and negative values, but the overall backlash, $\int_{-\infty}^{\infty} F_z(t)\,dt$, is along the positive $z$-axis, indicating that, on the whole, this component of the force tends to push the fiber up.

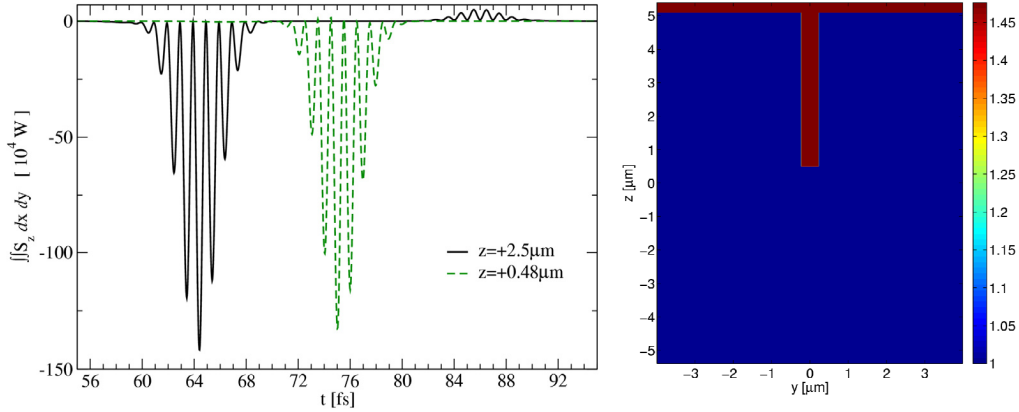

**Figure 5** (color online). Energy flux versus time far above the exit facet at $z_o=2.5\,\mu m$ (solid black), and just below the nano-filament's tip, at $z_o=0.48\,\mu m$ (dashed green). The reflected pulse, which carries ~4.7% of the incident pulse's energy, shows up in the solid black curve with ~21 fs delay caused by the round-trip to and from the exit facet.

The total electromagnetic momentum of the system at any instant of time is the integral of the Abraham momentum density $S(r,t)/c^2$ over the entire space, which includes the volume of the fiber, the vacuum surrounding the fiber, and the vacuum below the fiber, into which the hemi-spherical wave emerges. Figure 6 also shows (open circles, green) the time-derivative of the total Abraham momentum of the system, $-(d/dt)\iiint_{-\infty}^{\infty}[S_z(x,y,z,t)/c^2]dx\,dy\,dz$, which coincides precisely with the computed Lorentz force $F_z(t)$. The equality of $\boldsymbol{F}(t)$ and $-d\boldsymbol{p}_{EM}/dt$ is, of course, not accidental but, as discussed in [4], a general property of Maxwell's equations and the Lorentz law of force.

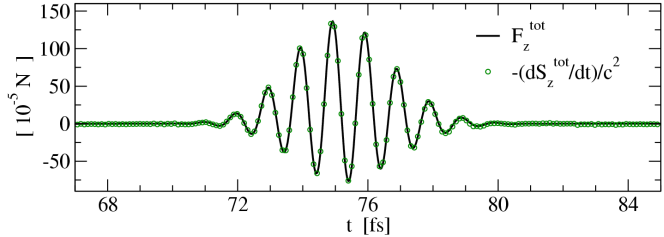

**Figure 6** (color online). Computed plot of $F_z(t)$, the total Lorentz force exerted on the fiber. The pulse reaches the tip at $t\approx 70$ fs and leaves the fiber by $t\approx 80$ fs. Overlapped with $F_z(t)$ and shown as open circles (green) is the time-derivative of the total EM momentum of the system [i.e., $S(r,t)/c^2$ integrated over the entire space], which coincides with the Lorentz force $F_z(t)$.

The time variations of the EM momentum contained in various regions of space are depicted in Fig. 7. The dashed (red) curve shows the integrated $S(r,t)/c^2$ over the upper half-space, $z>0.5\,\mu m$, which contains the nano-fiber and its surrounding free-space. Initially, the EM momentum in this half-space is large and negative ($-5.7\times10^{-18}$ N·s), as the light pulse propagates downward in the form of a guided mode of the filament. This is the Abraham momentum of the pulse inside the fiber, namely, $\mathcal{E}_{pulse}/(n_g c)$. Eventually, this momentum approaches zero as the light leaves the fiber, except for a small residual momentum after $t\approx 80$ fs, corresponding to the fraction of the incident pulse reflected at the exit facet. The reflected pulse has a positive EM momentum, as it propagates upward.

The dash-dotted (green) curve in Fig. 7 shows the integrated $S(r,t)/c^2$ over the lower half-space, $z<0.5\,\mu m$, into which the light pulse emerges from the fiber tip during the interval $t\sim 70-80$ fs, then proceeds to expand laterally. Initially, in the absence of any light, the momentum content of the lower half-space is zero. However, once the light emerges into this region, the electromagnetic momentum increases (in the negative z-direction) until it stabilizes at $t \sim 80$ fs, when the entire pulse has left the fiber. The EM momentum that emerges from the fiber is thus $p_z^{(EM)} = -7.1 \times 10^{-18}$ N·s. Given the pulse's energy inside the fiber ($\mathcal{E}_{pulse} = 2.78$ nJ) and the exit facet's transmissivity ($T = 1 - R = 95.3\%$), the naively-expected emergent momentum should be $p_z^{(EM)} = -T\mathcal{E}_{pulse}/c = -8.83 \times 10^{-18}$ N·s. The actual momentum, however, has only ~80% of this expected value, in fair agreement with the analysis of Section 2 for a point dipole.

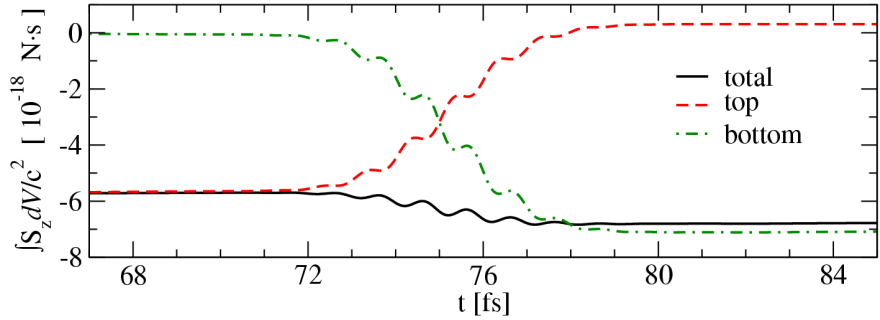

**Figure 7** (color online). Time variations of the EM momentum contained in various regions of space. The dashed (red) curve shows the integrated $S(r,t)/c^2$ over the upper half-space, $z > 0.5$ μm, which contains the filament and its surrounding free-space. The dash-dotted (green) curve shows the integrated $S(r,t)/c^2$ over the lower half-space, $z < 0.5$ μm, into which the light pulse emerges from the fiber tip during the interval $t \sim 70-80$ fs. The solid (black) curve, being the algebraic sum of the other two curves, represents the time variations of the total EM momentum of the system.

The solid (black) curve in Fig. 7, being the algebraic sum of the other two (red and green) curves, represents the time variations of the total EM momentum of the system. The net change of this EM momentum, obtained by subtracting the final value of the solid (black) curve from its initial value, is precisely equal to the time-integrated upward force, $\int_{-\infty}^{\infty} F_z(t) \, dt$, exerted on the fiber's end by the exiting light pulse. The backlash momentum (i.e., time-integrated force) is thus found to be $1.07 \times 10^{-18}$ N·s.

Although the backlash momentum obtained in our simulations is along the positive z-axis, one must be careful in interpreting it as the *only* momentum delivered to the filament when the light pulse passes from the fiber into the free space below. As argued earlier, the mechanical momentum of $-3.3 \times 10^{-18}$ N·s brought in by the light pulse is greater than the above backlash (even after subtracting the reflected $\boldsymbol{p}_{mech}$). Our conclusion, therefore, is that the total momentum delivered to the fiber tip is directed along the negative z-axis, exerting a net pull (rather than push) force on the fiber's end. In any event, what we have computed so far is the total contribution of the Lorentz force of the light pulse to the time-integrated force exerted on the fiber tip, a contribution that we have shown to be attributable to the changes in the electromagnetic as well as mechanical momenta of the light pulse. Since we have chosen an extremely short light pulse (duration ~10 fs) for these simulations, the Lorentz force delivered to the fiber tip has been impulsive, uncontaminated by the acoustic wave that the light pulse inevitably invokes in its wake. To determine the dynamic behavior of the nano-filament once the light pulse has departed, it is essential that one take into consideration the elastic properties of the fiber (including the tapered region that precedes the nano-filament), the role of the induced acoustic waves, and the interactions between the fiber and its mount.

**4. Concluding remarks**. The analysis of the preceding section indicates that the net force experienced by the fiber tip should be a push force if the mechanical momentum is ignored, but that it will be a pull force if $\boldsymbol{p}_{mech}$ is fully accounted for. We did not discuss the possibility of the mechanical momentum diffusing away from the light pulse (i.e., acoustic wave generation and propagation), as such an analysis would have required a detailed knowledge of the mechanical properties of the nano-filament as well as solution of the relevant elasticity equations coupled to the Maxwell-Lorentz equations. So long as the host medium is rigid enough to retain its opto-mechanical properties while absorbing the localized push and pull of the light pulse, while, at the same time, it is malleable enough for the acoustic waves to propagate at substantially below the speed of light, we believe the mechanical momentum of a short light pulse will remain more or less confined to within the boundaries of the pulse itself. Under such circumstances, it is reasonable to expect that the entire $\boldsymbol{p}_{mech}$ computed in the preceding section will be delivered to the nano-filament's tip upon arrival at the exit facet. This is not to say that the light pulse does not invoke an acoustic wave in its wake, considering that the nano-filament is preceded by a tapered region of the optical fiber, which, at some distance from the exit facet, is attached to a rigid mount. The invoked acoustic wave, however, arrives at the nano-filament's exit facet with some delay, at which time the Lorentz force of the exiting light pulse has already initiated the motion/deformation of the nano-filament.

The tug of the light pulse at the end face of the fiber sets in motion a sequence of events involving the propagation of an acoustic wave up the length of the fiber, reflection of this wave from the mount (to which the fiber is fastened), and the arrival of the reflected (acoustic) wave at the fiber's end face. These phenomena, however, being purely mechanical, can be analyzed with known mathematical and numerical methods.

Note that we are not arguing here in favor of either Abraham or Minkowski. Our analysis in the preceding sections indicates that the net force on the end face of the nano-filament must have contributions from electromagnetic as well as mechanical momenta of the light pulse. While the EM momentum is generally associated with the name of Abraham, the specific combination of $\boldsymbol{p}_{EM}$ and $\boldsymbol{p}_{mech}$ that is ultimately responsible for the observed force is attributable neither to Abraham nor to Minkowski.

It is our belief that measuring the push or pull force at the end face of a nano-filament, as proposed by She *et al* in [1], is quite possibly a valuable tool for investigating the momentum of light inside material media. The present paper has shown the feasibility of detailed theoretical calculation of the relevant optical forces in such experiments. What is now needed is accurate measurements in a well-controlled environment whose quantitative results could be compared with those of theoretical modeling and numerical simulations, before any definitive conclusions can be drawn with regard to the nature of optical momentum inside dielectric media.

**Appendix**

In general, the momentum $\boldsymbol{p}=p_z\hat{z}$ of a light pulse of energy $\mathcal{E}_{pulse}$, propagating in free space along the $z$-axis, satisfies the inequality $|\boldsymbol{p}|=p_z\leq \mathcal{E}_{pulse}/c$. Only when the spatial-frequency content of the pulse is confined to a small region of the $k$-space in the vicinity of $(k_x, k_y) = (0, 0)$, does $p_z$ approach its upper limit of $\mathcal{E}_{pulse}/c$. The momentum will be significantly lower than $\mathcal{E}_{pulse}/c$ if, at some point along its path, the light happens to be focused to a small diameter in the cross-sectional $xy$-plane. To prove the above statements, consider a finite-duration and finite-diameter light pulse whose $E$- and $H$-fields in the free space are expressed in terms of a plane-wave spectrum of spatio-temporal frequencies $(k_x, k_y, \omega)$, namely,

$$\boldsymbol{E}(\boldsymbol{r},t) = \tfrac{1}{2}\iiint \boldsymbol{\mathcal{E}}(k_x,k_y,\omega)\exp[\mathrm{i}(k_x x + k_y y + k_z z - \omega t)]\mathrm{d}k_x \mathrm{d}k_y \mathrm{d}\omega, \tag{A1a}$$

$$\boldsymbol{H}(\boldsymbol{r},t) = \tfrac{1}{2}\iiint \boldsymbol{\mathcal{H}}(k_x,k_y,\omega)\exp[\mathrm{i}(k_x x + k_y y + k_z z - \omega t)]\mathrm{d}k_x \mathrm{d}k_y \mathrm{d}\omega. \tag{A1b}$$

In general, $k_z = (\omega/c)\sqrt{1 - (ck_x/\omega)^2 - (ck_y/\omega)^2}$. For the fields to be real-valued it is necessary and sufficient that their Fourier transforms be Hermitian, that is,

$$\boldsymbol{\mathcal{E}}(k_x,k_y,\omega) = \boldsymbol{\mathcal{E}}^*(-k_x,-k_y,-\omega), \qquad \boldsymbol{\mathcal{H}}(k_x,k_y,\omega) = \boldsymbol{\mathcal{H}}^*(-k_x,-k_y,-\omega). \tag{A2}$$

Moreover, if the beam's cross-section in the $xy$-plane is required to be symmetric, say, with respect to the origin, that is, if the field amplitudes are to remain intact upon switching $(x,y)$ to $(-x,-y)$, then we must have

$$\boldsymbol{\mathcal{E}}(k_x,k_y,\omega) = \boldsymbol{\mathcal{E}}(-k_x,-k_y,\omega), \qquad \boldsymbol{\mathcal{H}}(k_x,k_y,\omega) = \boldsymbol{\mathcal{H}}(-k_x,-k_y,\omega). \tag{A3}$$

For the beam defined above, the Poynting vector may be written as follows:

$$\boldsymbol{S}(\boldsymbol{r},t) = \boldsymbol{E}(\boldsymbol{r},t)\times\boldsymbol{H}(\boldsymbol{r},t) = \tfrac{1}{4}\iiiint\!\!\!\int \boldsymbol{\mathcal{E}}(k_x,k_y,\omega)\times\boldsymbol{\mathcal{H}}(k'_x,k'_y,\omega')\exp[\mathrm{i}(k_x+k'_x)x]\exp[\mathrm{i}(k_y+k'_y)y]$$

$$\times \exp[\mathrm{i}(k_z+k'_z)z]\exp[-\mathrm{i}(\omega+\omega')t]\mathrm{d}k_x\mathrm{d}k_y\mathrm{d}\omega\mathrm{d}k'_x\mathrm{d}k'_y\mathrm{d}\omega'. \tag{A4}$$

Integrating $S_z(\boldsymbol{r},t)$ over the beam's cross-sectional area in the $xy$-plane and over all time, then using the identity

$$\int_{-\infty}^{\infty} \exp[\mathrm{i}(k+k')\zeta]\mathrm{d}\zeta = \delta(k+k'), \tag{A5}$$

where $\delta(k)$ is Dirac's delta function, the pulse's total energy content turns out to be

$$\mathcal{E}_{\mathrm{pulse}} = \iiint_{-\infty}^{\infty} S_z(x,y,z=z_0,t)\mathrm{d}x\mathrm{d}y\mathrm{d}t = \tfrac{1}{4}\iiint_{-\infty}^{\infty}[\boldsymbol{\mathcal{E}}(k_x,k_y,\omega)\times\boldsymbol{\mathcal{H}}(-k_x,-k_y,-\omega)]_z\mathrm{d}k_x\mathrm{d}k_y\mathrm{d}\omega$$

$$= \tfrac{1}{4}\iiint_{-\infty}^{\infty}[\boldsymbol{\mathcal{E}}(k_x,k_y,\omega)\times\boldsymbol{\mathcal{H}}^*(k_x,k_y,\omega)]_z\mathrm{d}k_x\mathrm{d}k_y\mathrm{d}\omega. \tag{A6}$$

To obtain the total momentum $\boldsymbol{p}$ of the light pulse, we integrate $\boldsymbol{S}(\boldsymbol{r},t)/c^2$ over the spatial coordinates $x$, $y$ and $z$. The result turns out to be independent of time, as follows:

$$\boldsymbol{p} = (1/c^2)\iiint_{-\infty}^{\infty} \boldsymbol{S}(\boldsymbol{r},t)\mathrm{d}x\mathrm{d}y\mathrm{d}z$$

$$= (4c^2)^{-1}\iiiint_{-\infty}^{\infty}\boldsymbol{\mathcal{E}}(k_x,k_y,\omega)\times\boldsymbol{\mathcal{H}}(-k_x,-k_y,\omega')\exp[-\mathrm{i}(\omega+\omega')t]\delta(k_z+k'_z)\mathrm{d}k_x\mathrm{d}k_y\mathrm{d}\omega\mathrm{d}\omega'$$

$$= (4c)^{-1}\iiint_{-\infty}^{\infty}\boldsymbol{\mathcal{E}}(k_x,k_y,\omega)\times\boldsymbol{\mathcal{H}}^*(k_x,k_y,\omega)\sqrt{1-(ck_x/\omega)^2-(ck_y/\omega)^2}\,\mathrm{d}k_x\mathrm{d}k_y\mathrm{d}\omega. \tag{A7}$$

In arriving at Eq. (A7) we have used the identity

$$\delta(k_z+k'_z) = \delta\!\left[(\omega/c)\sqrt{1-(ck_x/\omega)^2-(ck_y/\omega)^2} + (\omega'/c)\sqrt{1-(ck_x/\omega')^2-(ck_y/\omega')^2}\right]$$

$$= (\partial k_z/\partial\omega)^{-1}\delta(\omega+\omega') = c\sqrt{1-(ck_x/\omega)^2-(ck_y/\omega)^2}\,\delta(\omega+\omega'). \tag{A8}$$

The coefficient $\sqrt{1-(ck_x/\omega)^2-(ck_y/\omega)^2}$ appearing in the integrand in Eq. (A7) is simply the obliquity factor $ck_z/\omega = k_z/k_0 = \cos\theta$, where $\theta$ is the deviation angle of the $k$-vector from the $z$-axis. Comparing Eq. (A6) with Eq. (A7) makes it clear that the relation between the field energy and momentum will approach $p_z \approx \mathcal{E}_{\text{pulse}}/c$ only when $\sqrt{k_x^2 + k_y^2} \ll \omega/c$, namely, when the beam's cross-sectional diameter is substantially wider than a wavelength. The $k$-space spectrum broadens, however, as the beam diameter shrinks, resulting in a substantial reduction of $p_z$ below its peak value of $\mathcal{E}_{\text{pulse}}/c$.

**Acknowledgements.** This work has been supported by the Air Force Office of Scientific Research (AFOSR) under contract number FA 9550−04−1−0213. The authors are grateful to Andrey Kobyakov for helpful discussions.